\documentclass[10pt,twocolumn,twoside]{IEEEtran}
\usepackage{latexsym,amsmath,amssymb,amsfonts,mathrsfs,amsbsy}
\usepackage[dvips]{graphicx}
\usepackage[final]{epsfig}
\usepackage{cases}
\usepackage{url}
\usepackage{multirow}
\usepackage{array,colortbl}
\usepackage{color}
\usepackage{textpos}
\usepackage{hyperref}
\usepackage{algorithm}
\usepackage{algorithmic}
\usepackage{array}
\usepackage{amssymb}
\usepackage{times}
\usepackage{setspace}
\usepackage{cite}
\usepackage{epstopdf} 
\usepackage{tabularx,ragged2e,booktabs,caption}
\usepackage{mathrsfs}
\usepackage{amsmath}
\usepackage{arydshln}
\newcommand{\fontOfEquations}{}
\newcommand{\vsapceOfFigures}{\vspace{-0.3cm}}
\newcommand{\FigureSize}{0.90}
\newcommand{\IfNewPage}{\newpage}
\begin{document}
\title{Cram\'er Rao Lower Bound for Underwater Range Estimation with Noisy Sound Speed Profile}

%\small
\normalsize

% author names and affiliations
\author{Hamid Ramezani*, Raj Thilak Rajan, ~\IEEEmembership{Student Member,~IEEE},
        and Geert Leus,~\IEEEmembership{Fellow,~IEEE}
        
\thanks{
H. Ramezani and G.Leus are with the Faculty of Electrical Engineering, Mathematics and Computer Science, Delft University of Technology, Delft, The Netherlands. (e-mail: h.mashhadiramezani@tudelft.nl and g.j.t.leus@tudelft.nl) 

R. T. Rajan is with the Faculty of Electrical Engineering, Mathematics and Computer Science, Delft University of Technology, The Netherlands, and also with Netherlands Institute for Radio Astronomy (ASTRON), Dwingeloo, The Netherlands. (e-mail: rajan@astron.nl)
%The authors are with the Faculty of Electrical Engineering, Mathematics and Computer Science, Delft University of Technology, 2826 CD Delft, The Netherlands. e-mails: h.mashhadiramezani@tudelft.nl, rajan@astron.nl, and g.j.t.leus@tudelft.nl.
}

%\thanks{* Corresponding author: Hamid Ramezani, phone: (+31)152786280, fax: (+31)152786190, e-mail: h.mashhadiramezani@tudelft.nl.}
\thanks{The research leading to these results has received funding in part from the European Commission FP7-ICT Cognitive Systems, Interaction, and Robotics under the contract \#270180 (NOPTILUS)}
}
\markboth{IEEE Signal Processing Letter (submitted)}{}
\maketitle
%====================================================================
%-------------------------------------------------------------
\begin{abstract}
In this paper, the Cram\'er Rao bound (CRB) for range estimation between two underwater nodes is calculated under a Gaussian noise assumption on the measurements. The nodes can measure their depths, their mutual time of flight, and they have access to noisy sound speed samples at different depths. The effect of each measurement on the CRB will be analyzed, and it will be shown that for long distances, the effect of the sound speed measurement noise is dominant, and its impact depends on the positions of the nodes, actual sound speed profile, the number of sound speed samples, and the depths at which the sound speed samples are gathered.  
\end{abstract}
%==============================================================
\IEEEpeerreviewmaketitle
%==============================================================
\section{Introduction}
Range estimation is required mostly for sensor network localization and navigation. In this paper, we present an insight into underwater ranging via the Cram\'er Rao bound (CRB). The CRB expresses a lower bound on the variance of unbiased estimators of a deterministic parameter. 
The CRB for range estimation in a terrestrial environment where sensor nodes communicate with each other through radio frequency links has been investigated in \cite{cardinali2006uwb}. There, time of flight (ToF) measurements are used for range estimation, and it is shown that the corresponding CRB depends on the signal bandwidth, wave propagation speed, and received signal to noise ratio. Apart from range information in the time delay, \cite{wang2009cramer} has reached a more accurate formulation by also extracting range information from the amplitude of the received signal power. Although the results of the above papers give us a valid inception of the range estimation accuracy in free space, they do not justify why practical underwater range estimation (specifically for long distances) suffers from a higher inaccuracy than anticipated by the developed bounds.

Acoustic underwater communications is quite different from its terrestrial counterparts \cite{stojanovic2009underwater}. The propagation speed is not constant and it varies with temperature, salinity and pressure \cite{coppens1981simple}. On the other hand, the underwater sensor nodes have the privilege to measure their depth via a pressure sensor. In \cite{ramezani2012ranging}, it is shown that knowing the depth information and the sound speed profile (SSP), a mutual distance between two nodes can be obtained from a single ToF measurement. Under these conditions, the CRB for range estimation has been derived for a multiple-isogradient sound speed profile in \cite{ramezani2012ranging}, and for a more general sound speed profile in \cite{berger2008stratification}. However, in practice, the SSP has to be measured. Consequently, a noisy sound speed measurement will indirectly affect the accuracy of range estimation. 

In general, the SSP can be represented as a linear combination of $N$ basis functions obtained from empirical data as \cite{carriere2007dynamic}
\begin{equation}
\fontOfEquations
c(z) = \bar{c}(z)+\sum_{n = 1}^{N}{a_nf_n(z)}, \label{eq:sspBasis}
\end{equation}
where $z$ represents depth, $\bar{c}(z)$ is the nominal sound speed profile which is known a priori (obtained from historical data), and $f_n(z)$ for $n=1,2,...,N$ denotes the basis functions. 
In order to measure the sound speed at a certain depth, a CTD (conductivity, temperature, and depth) sensor is used. Gathering all the noisy sound speed measurements at $M$ different depths leads to
\begin{equation}
\fontOfEquations
\hat{\bf c} = \bar{\bf c}+{\bf F}{\bf a}+{\bf v},
\end{equation}
where $\hat{\bf c} = [\hat{c}(z_1),\hat{c}(z_2),\hdots,\hat{c}(z_M)]^T$ is a vector of noisy sound speed samples at different depths, $\bar{\bf c} = [\bar{c}(z_1),\bar{c}(z_2),\hdots,\bar{c}(z_M)]^T$,  ${\bf a}=[a_1,a_2,...,a_N]^T$, ${\bf F}$ is  an $M\times N$ matrix with $n$-th column ${\bf f}_n = [f_n(z_1),f_n(z_2),\hdots,f_n(z_M)]^T,\ \forall\  n = 1,2,..,N$, and ${\bf v}$ is the zero-mean Gaussian distributed measurement noise with covariance ${\bf R}_{\bf v}=\sigma^2_c{\bf I}_{M\times M}$.

\section{Ray Tracing}
\begin{figure}
\centering
\includegraphics[width = \FigureSize \linewidth]{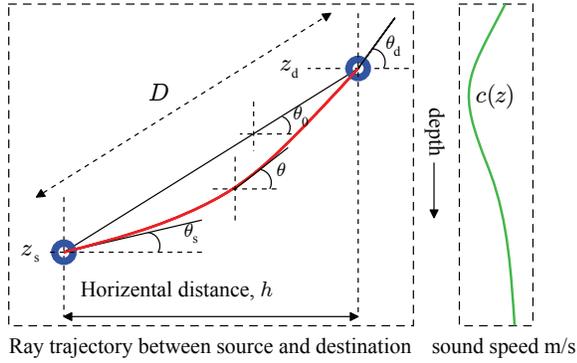} % 0.99
\caption{Ray propagation between the source and destination.}
\label{fig:letter_nodes}
%\vsapceOfFigures
\end{figure}

The relation between the ToF and the node positions can be extracted from a set of differential equations characterized by Snell's law:
\begin{equation}
\fontOfEquations
\frac{\cos \theta_\text{s}}{c(z_\text{s})}=\frac{\cos \theta_\text{d}}{c(z_\text{d})}=\frac{\cos \theta}{c(z)}=k_0,~~\theta \in \left(-\frac{\pi}{2},\frac{\pi}{2}\right),\label{eq:snell}
\end{equation}
where $k_0$ is constant for a given ray, $\theta_\text{s}$ and $\theta_\text{d}$ are the ray angle at the source and the destination, respectively, $\theta$ is the ray angle at any point between the source and the destination, and $z_\text{s}$ and $z_\text{d}$ are the depths of the source and destination, respectively, as shown in Fig.~\ref{fig:letter_nodes}.
The basic relationship between the ToF $t$, horizontal distance $h$, and depth $z$ can be represented by
\begin{align}
\fontOfEquations
\partial h &= \frac{\partial z}{\sin \theta},\nonumber\\ 
\partial t &= \frac{\partial h}{c(z)},\label{eq:diff}
\end{align}
which will be used in calculating the ToF and horizontal distance between two points. 

Although in an underwater medium with a general SSP, a ray between two points can have different patterns \cite{ramezani2012ranging}, here we assume that the ray crosses any depth between the two points only once. Using this assumption, the 
ToF and the horizontal distance can be formulated as (see \eqref{eq:snell} and \eqref{eq:diff})
\begin{align}
\fontOfEquations
t = & \int_{z_\text{s}}^{z_\text{d}}{\frac{1}{c(z)\sqrt{1-[k_0c(z)]^2}}d_z}, \label{eq:tof} \\
h = & \int_{z_\text{s}}^{z_\text{d}}{\frac{k_0c(z)}{\sqrt{1-[k_0c(z)]^2}}d_z}. \label{eq:hdist}
\end{align}

With the knowledge of the ToF, the SSP, and the depths, one can calculate $k_0$ from \eqref{eq:tof}, and use that in \eqref{eq:hdist} to find the horizontal distance and eventually the range $D$ between the two points. Unfortunately, the measurements are always noisy and that makes the estimation inaccurate. In the next section, we investigate the lowest achievable bound by any unbiased range estimator.

\section{Cram\'er Rao Lower Bound}
As explained before, the measurements are the ToF $t$, the depth of the source $z_\text{s}$, the depth of the destination $z_\text{d}$, and the samples of the sound speed at different depths $\hat{\bf c}$, each contaminated by respective Gaussian-distributed noise with zero mean and variance $\sigma_t^2$, $\sigma_z^2$, $\sigma_z^2$, and covariance $\sigma^2_c{\bf I}_{M\times M}$, respectively. Stacking all the measurements in a vector we have
\begin{equation}
{\bf f}^T = [t, z_\text{s}, z_\text{d}, \hat{{\bf c}}^T]_{1 \times (M+3)}, \nonumber
\end{equation}
with related noise ${\bf w}$ whose elements are assumed to be independent of each other, and therefore 
\begin{equation}
{\bf R}_{w}=\mathbb{E}[{\bf w}{\bf w}^T]=\text{diag}([\sigma_t^2, \sigma_z^2, \sigma_z^2, \sigma^2_c{\bf 1}_{1\times M}]). \nonumber
\end{equation}
The estimated parameters are stated in 
\begin{equation}
{\bf x}^T = [k_0,z_\text{s}, z_\text{d},{\bf a}^T]_{1\times (N+3)}. \nonumber
\end{equation}
Later, with a change of variables, the CRB for the horizontal distance and the mutual distance can be obtained. 

\IfNewPage
For Gaussian distributed noise, the elements of the Fisher information matrix (FIM) can be obtained as \cite{kay1998fundamentals} 
\begin{equation}
\fontOfEquations
[{\bf I}_{\bf x}]_{i,j} = 
\frac{\partial{\bf f}}{\partial{x_i}}^T
{\bf R}_w^{-1}
\frac{\partial{\bf f}}{\partial{x_j}}  
+
\frac{1}{2}\text{tr}
\left[{\bf R}_w^{-1}\frac{\partial{\bf R}_w}{\partial{x_i}}{\bf R}_w^{-1} \frac{\partial{\bf R}_w}{\partial{x_j}}\right]. \label{eq:FIM}
\end{equation}
Among the list of measurements, the variance of the ToF is distance dependent \cite{ramezani2013target}, and hence the second term of \eqref{eq:FIM} is not zero for $[{\bf I}_{\bf x}]_{1,1}$. However, it can be ignored for high values of the SNR. The diagonal elements of the inverse of the Fisher information matrix give us the lowest bound on the variance of any unbiased estimator for ${\bf x}$. The elements of the inverse FIM (which is symmetric) can be calculated as (see Appendix A)
\begin{align}
\fontOfEquations
[{\bf I}^{-1}_{\bf x}]_{1,1} =&
\sigma_t^2\frac{1}{\left(\partial t/\partial k_0\right)^2}+ \nonumber\\ &\sigma_z^2\frac{\left(\partial t/\partial z_\text{s}\right)^2}{\left(\partial t/\partial k_0\right)^2}+ \sigma_z^2\frac{\left(\partial t/\partial z_\text{d}\right)^2}{\left(\partial t/\partial k_0\right)^2} + \nonumber\\
&\sigma_c^2\frac{1}{\left(\partial t/\partial k_0\right)^2}\frac{\partial t}{\partial {\bf a}}\left({\bf F}^T{\bf F}\right)^{-1}\left[\frac{\partial t}{\partial {\bf a}}\right]^T,
 \nonumber
\end{align}
\begin{align}
\fontOfEquations
[{\bf I}^{-1}_{\bf x}]&_{1,2} =- \sigma_z^2\frac{\partial t/\partial z_\text{s}}{\partial t/\partial k_0},\nonumber\\
[{\bf I}^{-1}_{\bf x}]&_{1,3} =- \sigma_z^2\frac{\partial t/\partial z_\text{d}}{\partial t/\partial k_0},\nonumber\\
[{\bf I}^{-1}_{\bf x}]&_{1,4:N+3} = -\sigma^2_c\frac{1}{\partial t / \partial k_0} \frac{\partial t}{\partial {\bf a}}\left(F^TF\right)^{-1}, \nonumber \\
[{\bf I}^{-1}_{\bf x}]&_{2,2} = \sigma_z^2,\nonumber\\
[{\bf I}^{-1}_{\bf x}]&_{3,3} = \sigma_z^2,\nonumber\\
[{\bf I}^{-1}_{\bf x}]&_{2:4,4:N+3} = {\bf 0}, \nonumber \\
[{\bf I}^{-1}_{\bf x}]&_{4:N+3,4:N+3} = \sigma^2_c\left({\bf F}^T{\bf F}\right)^{-1}, \label{eq:IxInvElements}
\end{align}
where 
\begin{align}
\fontOfEquations
\frac{\partial t}{\partial k_0} = & \int_{z_\text{s}}^{z_\text{d}}\frac{k_0c(z)}{({1-[k_0c(z)]^2})^{\frac{3}{2}}}dz, \nonumber \\
\frac{\partial t}{\partial z_\text{s}} = & \frac{-1}{c(z_\text{s})\sqrt{1-[k_0c(z_\text{s})]^2}}, \nonumber \\
\frac{\partial t}{\partial z_\text{d}} = & \frac{1}{c(z_\text{d})\sqrt{1-[k_0c(z_\text{d})]^2}}, \nonumber
\end{align}
and $\frac{\partial t}{\partial {\bf a}}$ is a $1\times N$ vector $[\frac{\partial t}{\partial a_1},\frac{\partial t}{\partial a_2},...,\frac{\partial t}{\partial a_N}]$ including the derivatives of the ToF to the coefficients of the basis functions in \eqref{eq:sspBasis} and can be obtained as
\begin{equation}
\fontOfEquations
\frac{\partial t}{\partial a_n} = \int_{z_\text{s}}^{z_\text{d}}\frac{2[k_0c(z)]^2-1}{c^2(z)({1-[k_0c(z)]^2})^{\frac{3}{2}}}f_n(z)dz. \nonumber
\end{equation}

In 3D underwater  localization \cite{isik2009three}, it is shown that knowing the depths of the sensor nodes, only the horizontal distance between each pair of nodes can be used for self-localization. Here, our parameters of interest are ${\bf y}^T = [h,z_\text{s},z_\text{d},{\bf a}^T]$ whose CRB is given by a transform of \eqref{eq:IxInvElements} as 
\begin{equation}
\fontOfEquations
{\bf I}^{-1}_{\bf y} = {\bf H}^T{\bf I}_{\bf x}^{-1}{\bf H}, \label{eq:FIM_y1}
\end{equation}
where ${\bf H}$ is the Jacobian matrix of $\bf y$ with respect to $\bf x$ and can be formulated as
\begin{equation}
\fontOfEquations
{\bf H} = \left[\begin{array}{c|ccc}
\frac{\partial h}{\partial k_0} & 
\frac{\partial h}{\partial z_\text{s}} & 
\frac{\partial h}{\partial z_\text{d}} & 
\frac{\partial h}{\partial {\bf a}^T} \\   
          {\bf 0}_{(N+2)\times 1}& & {\bf I}_{(N+2)\times (N+2)} &\end{array}\right]^T, \label{eq:FIM_y}
\end{equation}
where
\begin{align}
\fontOfEquations
\frac{\partial h}{\partial z_\text{d}} =&  {k_0c^2(z_\text{d})}\frac{\partial t}{\partial z_\text{d}},
\nonumber\\
\frac{\partial h}{\partial { a_n}} =&  \int_{z_\text{s}}^{z_\text{d}}\frac{k_0}{({1-[k_0c(z)]^2})^{\frac{3}{2}}}f_n(z)dz 
\nonumber \\
\frac{\partial h}{\partial z_\text{s}} =&  {k_0c^2(z_\text{s})}\frac{\partial t}{\partial z_\text{s}},
\nonumber\\
\frac{\partial h}{\partial k_0} =&  \frac{1}{k_0}\frac{\partial t}{\partial k_0}, 
\end{align}
and ${\bf I}$ is the identity matrix. Using \eqref{eq:FIM_y} in \eqref{eq:FIM_y1}, and computing $[{\bf I}^{-1}_{\bf y}]_{11}$ as the CRB of $h$ results in
\begin{align}
\fontOfEquations
\text{CRB}_h =
&\sigma_t^2\frac{1}{k_0^2}+ \nonumber\\
&\sigma_z^2\frac{1-[k_0c(z_\text{s})]^2}{[k_0c(z_\text{s})]^2} +   
 \sigma_z^2\frac{1-[k_0c(z_\text{d}e)]^2}{[k_0c(z_\text{d})]^2} + \nonumber\\
&\sigma_c^2\|\frac{\partial h}{\partial {\bf a}}-\frac{1}{k_0}\frac{\partial t}{\partial {\bf a}}\|^2_{\left({\bf F}^T{\bf F}\right)^{-1}}
 \label{eq:CRB_h}
\end{align}
where $\|{\bf x}\|^2_A={\bf x}^TA{\bf x}$, and $\frac{\partial h}{\partial {\bf a}}-\frac{1}{k_0}\frac{\partial t}{\partial {\bf a}}$ can be viewed as the inner product of $g(z)$ and the basis functions for $z\in[z_\text{s},z_\text{d}]$ where 
\begin{equation}
\fontOfEquations
g(z) = \frac{1}{k_0c^2(z)\sqrt{1-[k_0c(z)]^2}}. \label{eq:g_z}
\end{equation} 

%In 3D underwater localization \cite{isik2009three}, it is shown that only the horizontal distance between the each pair of nodes can be used for self-localization. 

Assume that we have sampled the sound-speed profile linearly for $z\in[z_\text{s},z_\text{d}]$. It can then be shown that for a large number of samples ($M\rightarrow \infty$) we have 
\begin{equation}
\fontOfEquations
\|\frac{\partial h}{\partial {\bf a}}-\frac{1}{k_0}\frac{\partial t}{\partial {\bf a}}\|^2_{({\bf F}^T{\bf F})^{-1}} \approx (\Delta z{\bf g}^T{\bf F})({\bf F}^T{\bf F})^{-1}({{\bf F}}^T{\bf g}\Delta z) \label{eq:expansion}
\end{equation}
where $\Delta z=\frac{|z_\text{d}-z_\text{s}|}{M}$, and ${\bf g}$ is a $M\times 1$ vector whose elements are the samples of $g(z)$ at different depths. Since ${\bf F}\left({\bf F}^T{\bf F}\right)^{-1}{\bf F}^T$
can be seen as the projection matrix on the columns of ${\bf F}$, \eqref{eq:expansion} can be represented by 
\begin{equation}
\fontOfEquations
\|\frac{\partial h}{\partial {\bf a}}-\frac{1}{k_0}\frac{\partial t}{\partial {\bf a}}\|^2_{({\bf F}^T{\bf F})^{-1}} 
\approx 
\frac{|z_\text{d}-z_\text{s}|}{M}\int_{z_\text{s}}^{z_\text{d}}{\tilde{g}^2(z)dz}, \label{eq:expansion3}
\end{equation}
where $\tilde{g}(z)$ is the projection of $g(z)$ on the space which is spanned by the sound-speed basis functions which are defined in the range $[z_\text{s},z_\text{d}]$. Since the energy of the signal is always greater than or equal to the energy of its projection, the right hand side of \eqref{eq:expansion3} can be approximately upper-bounded by 
$\frac{|z_\text{d}-z_\text{s}|}{M}E_g$ 
where $E_g$ is the energy of $g(z)$ within the range $[z_\text{s},z_\text{d}]$.
Note that as $M$ increases the effect of a noisy sound speed measurement on the CRB in \eqref{eq:CRB_h} decreases.

The CRB of the range (denoted by $D$) estimation as a function of $h$, $z_\text{s}$, and $z_\text{d}$, i.e., $D=\sqrt{h^2+(z_\text{s}-z_\text{d})^2}$, can now be formulated as
\begin{equation}
\text{CRB}_D = {\bf s}^T~[{\bf I}^{-1}_{\bf y}]_{1:3,1:3}~{\bf s}, \label{eq:CRB_D}
\end{equation}
where ${\bf s}=\partial D/\partial[h,z_\text{s},z_\text{d}] =  [\cos\theta_0,~~-\sin\theta_0,~~\sin\theta_0]^T$,
and $\theta_0$ is the angle between the straight line from the source to the destination and the horizontal axis.
The CRB of $D$ can be simplified as 
\begin{align}
\fontOfEquations
\text{CRB}_D = 
&\sigma_t^2c(z_\text{s})^2\left(\frac{\cos\theta_0}{\cos\theta_\text{s}}\right)^2+ \nonumber\\
&\sigma_z^2\left(\frac{\sin[\theta_0-\theta_\text{d}]}{\cos\theta_\text{d}}\right)^2 + \sigma_z^2\left(\frac{\sin[\theta_0-\theta_\text{s}]}{\cos\theta_\text{s}}\right)^2+ 
\nonumber\\
&\sigma_c^2\left(\cos\theta_0\right)^2\|\frac{\partial h}{\partial {\bf a}}-\frac{1}{k_0}\frac{\partial t}{\partial {\bf a}}\|^2_{\left({\bf F}^T{\bf F}\right)^{-1}}, \label{eq:CRB_D_Final}
\end{align}
where the first two terms (related to a noisy ToF and depth measurement at the destination) are similar to what is extracted in \cite{berger2008stratification}. Regarding the noisy sound speed samples, it can be observed that five factors affect the CRB: the measurement noise power, how the ray propagates (the actual sound speed profile), the number of samples $M$, the depth at which the samples are taken, and the inner product of $g(z)$ with the truncated form of the basis functions.
%It can be proven that if we have a set of basis functions within depths $[z_\text{s},z_\text{d}]$ that $g(z)$ can be expanded without any residual, then the forth term in \eqref{eq:CRB_D_Final} is independent of the basis functions for large sound speed samples obtained linearly at $[z_\text{s},z_\text{d}]$, however the upper-bounded is
%\begin{equation}
%\fontOfEquations
%\|\frac{\partial h}{\partial {\bf a}}-\frac{1}{k_0}\frac{\partial t}{\partial {\bf a}}\|^2_{({\bf F}^T{\bf F})^{-1}} \lesssim \frac{|z_\text{d}-z_\text{s}|}{M}\int_{z_\text{s}}^{z_\text{d}}{g^2(z)dz}.
%\end{equation}

The effects of noisy depth and ToF measurements on the range estimation have been analyzed before in \cite{berger2008stratification} and \cite{ramezani2012ranging} for a known SSP. In the numerical section, we focus more on the effect of noisy sound speed samples.

%The equation \eqref{eq:CRB_D} can be approximated by 
%\begin{equation}
%\text{CRB}_D > \text{CRB}_h\cos^2\theta_0 + %2\sigma^2_z\sin^2\theta_0, \label{eq:CRB_Dapp}
%\end{equation}
%which means the depth information from time of flight %measurement has been ignored. In the numerical section we %will show how close this approximation is to its actual %value.

\begin{figure}[b]
\centering
\includegraphics[width=\FigureSize \linewidth]{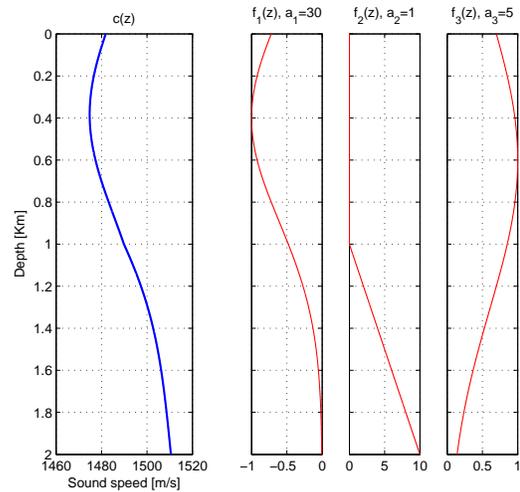}
\caption{The sound speed profile as a function of depth is presented, along with corresponding coefficients of each basis function.}
\label{fig:letter_SSP}
\vsapceOfFigures
\end{figure}

\section{Numerical Results}
In this section, we evaluate the CRB of the range estimation for a given set up. A 2D environment with length $D_h = 10$Km and depth $D_z=2$Km is considered. The nominal sound speed is set to $\bar{c}(z) = 1500$m/s, and it is assumed that the SSP is composed of three basis functions as depicted in Fig.~\ref{fig:letter_SSP}. The depth of the source node is set to $z_\text{s}=0$, and the coordinate of the destination point $[h,z_\text{d}]$ varies over the area where the path trajectory crosses any depth only once. In addition, there are $M=10$ sound speed samples obtained at depth $z_m=m\frac{D_z}{M}$ for $m=\{1,2,...,M\}$. The variance of the measurements are $\sigma^2_t=10^{-8}$, $\sigma^2_z = 1$, and $\sigma^2_c = 1$.

\begin{figure}[H]
\centering
\includegraphics[width=\FigureSize\linewidth]{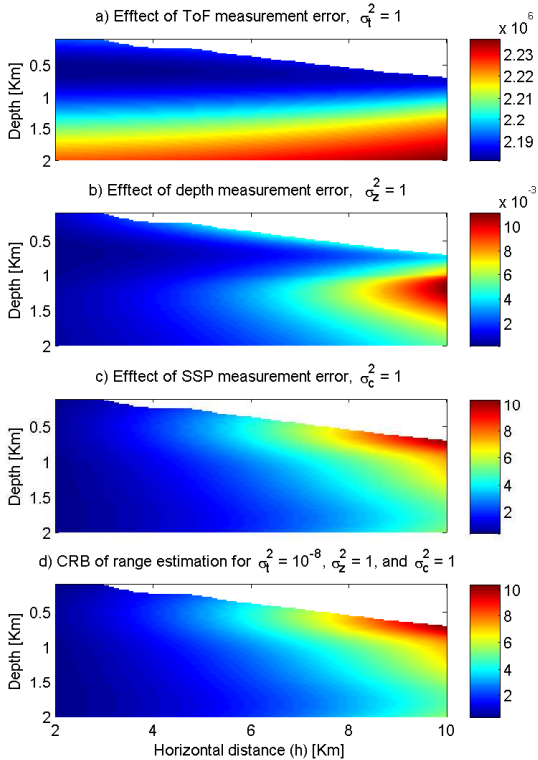}
\caption{{CRB of range estimation, a) effect of noisy ToF measurement, b) effect of noisy depth measurement, c) effect of noisy sound speed sample, d) the overall CRB of range estimation}.}
\label{fig:letter_D_CRB}
\vsapceOfFigures
\end{figure}

Fig.~\ref{fig:letter_D_CRB} illustrates the effect of each phenomenon on the CRB of $D$ in part a, b, and c for a normalized noise power, and the overall CRB for $D$ in part d. The white rejoins in these figures indicate that the initial assumption of path trajectory does not hold. 
It can be observed that for long distances a noisy ToF measurement deteriorates the performance almost similarly for any position of the destination point, and for a small ToF measurement noise power, its effect is trivial. Furthermore, the depth measurement error is not influential on the CRB for actual values of the noise power (e.g., $\sigma^2_z = 1$). In contrast, the effect of a noisy SSP is dominant here, and it is more dominant when the vertical distance between the source and the destination is lower than the horizontal distance.

%\begin{figure}
%\centering
%\includegraphics[width=0.90\linewidth]{./letter_L_Err.eps}
%\caption{}
%\label{fig:letter_L_Err}
%\end{figure}

\section{Conclusions}
The CRB of range estimation in an underwater sensor network has been derived under a depth-dependent sound-speed profile wherein noisy time of flight, depth, and sound speed measurements are available. The effect of each measurement noise on the CRB of range estimation has been evaluated analytically. For long distances, the noise power of the depth measurements does not play a significant role in the CRB, while those of the ToF and the sound speed samples are dominant. Over long distances, even with perfect ToF measurements, the range estimation cannot be perfect. We have shown that for a few sound speed samples at different depths, several factors play a vital role in the CRB such as the basis functions that the sound speed profile (SSP) is composed of, the actual SSP, the number of sound speed samples, and the positions of the source and the destination.

\IfNewPage
\section{Appendix}
The Fisher information matrix can be represented as
\begin{equation}
{\bf I}_{\bf x} = \left[\begin{matrix}{\bf A} & {\bf B} \\ {\bf B}^T & {\bf D}\end{matrix}\right] \label{eq:FIMblock}
\end{equation}
where
\begin{align}
{\bf A} = & \frac{1}{\sigma_t^2}\left[\begin{matrix}
\frac{\partial t}{\partial k_0} \\ \frac{\partial t}{\partial z_s} \\ \frac{\partial t}{\partial z_\text{d}} \end{matrix}\right]\left[\begin{matrix}
\frac{\partial t}{\partial k_0} & \frac{\partial t}{\partial z_s} & \frac{\partial t}{\partial z_\text{d}} \end{matrix}\right]+\frac{1}{\sigma_z^2}\left[\begin{matrix}
0 & 0 & 0 \\
0 & 1 & 0 \\
0 & 0 & 1
\end{matrix}\right],\nonumber \\
{\bf B} =& \frac{1}{\sigma_t^2}\left[\begin{matrix}
\frac{\partial t}{\partial k_0}, \frac{\partial t}{\partial z_s}, \frac{\partial t}{\partial z_\text{d}} \end{matrix}\right]^T
\frac{\partial t}{\partial {\bf a}}, \nonumber \\
{\bf D} =& \frac{1}{\sigma^2_t}\left[\frac{\partial t}{\partial {\bf a}}\right]^T\frac{\partial t}{\partial {\bf a}} + \frac{1}{\sigma_c^2}{\bf F}^T{\bf F},
\end{align}
and ${\bf A}$ is a $3\times 3$ symmetric positive definite, ${\bf B}$ is a $3 \times N$, and ${\bf D}$ is a $N\times N$ matrices. 

Using the general formula of matrix inversion in block form, the inverse FIM in \eqref{eq:FIMblock} can be obtained as
\begin{align}
&\left[{\bf I}_{\bf x}^{-1}\right] = \nonumber\\
&\footnotesize{\left[\begin{matrix}
({\bf A}-{\bf B}{\bf D}^{-1}{\bf B}^T)^{-1} &
-{\bf A}^{-1}{\bf B}({\bf D}-{\bf B}^T{\bf A}^{-1}{\bf B})^{-1} \\
-({\bf D}-{\bf B}^T{\bf A}^{-1}{\bf B})^{-1}{\bf B}^T{\bf A}^{-1} &
 ({\bf D}-{\bf B}^T{\bf A}^{-1}{\bf B})^{-1} \end{matrix}\right]}.   \label{eq:MatInvBlock}    
\end{align}
The first block of \eqref{eq:MatInvBlock} can be further expanded according to the Woodbury identity as \cite{golub2012matrix}
\begin{align}
({\bf A}- & {\bf B}{\bf D}^{-1}{\bf B}^T)^{-1} = \nonumber \\ 
& {\bf A}^{-1}+{\bf A}^{-1}{\bf B}({\bf D}-{\bf B}^T{\bf A}^{-1}{\bf B})^{-1}{\bf B}^T{\bf A}^{-1}, \label{eq:woodbury}
\end{align}
where the elements of the ${\bf A}^{-1}$ are
\begin{align}
\hspace{-1.8cm}
[{\bf A}^{-1}]_{11} = & \frac{\sigma_t^2}{\left(\partial t/\partial k_0\right)^2}+  \sigma_z^2\frac{\left(\partial t/\partial z_\text{s}\right)^2}{\left(\partial t/\partial k_0\right)^2}+  \sigma_z^2\frac{\left(\partial t/\partial z_\text{d}\right)^2}{\left(\partial t/\partial k_0\right)^2} \nonumber \\
[{\bf A}^{-1}]_{12} = & -\sigma_z^2\frac{\partial t/\partial z_\text{s}}{\partial t/\partial k_0} \nonumber \\
[{\bf A}^{-1}]_{13} = &-\sigma_z^2\frac{\partial t/\partial z_\text{d}}{\partial t/\partial k_0} \nonumber \\
[{\bf A}^{-1}]_{23}= & 0,~~~~
[{\bf A}^{-1}]_{22}= \sigma_z^2. \label{eq:AInv}
\end{align}
Using \eqref{eq:AInv} in ${\bf A}^{-1}{\bf B}$ leads to
\begin{align}
{\bf A}^{-1}{\bf B} = \left[\begin{matrix}
\frac{1}{\partial t / \partial k_0} \\
0 \\
0
\end{matrix}\right]\frac{\partial t}{\partial {\bf a}} \nonumber \\
{\bf B}^T{\bf A}^{-1}{\bf B} = \frac{1}{\sigma^2_t}\left[\frac{\partial t}{\partial {\bf a}}\right]^T\frac{\partial t}{\partial {\bf a}},
\end{align}
therefore the second term of \eqref{eq:woodbury} is
\begin{align}
{\bf A}^{-1} & {\bf B}({\bf D}-{\bf B}^T{\bf A}^{-1}{\bf B})^{-1}{\bf B}^T{\bf A}^{-1}= \nonumber\\ &\frac{1}{\left(\partial t / \partial k_0\right)^2} \left[ \begin{matrix}
\frac{\partial t}{\partial {\bf a}}\left(\frac{1}{\sigma^2_c}{\bf F}^T{\bf F}\right)^{-1}\left[\frac{\partial t}{\partial {\bf a}}\right]^T & 0 & 0 \\
0 & 0 & 0 \\
0 & 0 & 0 
\end{matrix}
\right],
\end{align}
which means that the second term of \eqref{eq:woodbury} only affects the lower bound of the variance of the $k_0$ estimator. Using a similar approach for the other matrix blocks in \eqref{eq:MatInvBlock} we obtain \eqref{eq:IxInvElements}.

\IfNewPage
\bibliographystyle{IEEEtran}
\small
\bibliography{Letter01}

\end{document}